\pdfoutput=1
\documentclass[a4paper,11pt]{article}
\usepackage{pos}

\title{Measurements of electroweak penguin $B$ decays with missing energy at Belle and Belle II}

\author*[a]{Lorenz G\"artner}

\onbehalf{on behalf of the Belle II Collaboration}

\affiliation[a]{Faculty of Physics, Ludwig-Maximilians-Universit\"at M\"unchen,\\
  Munich, Germany}

\emailAdd{lorenz.gaertner@lmu.de}

\abstract{
  Electroweak penguin $B$ decays with neutrinos or $\tau$ leptons in the final state are sensitive probes of physics beyond the Standard Model. Using the full Belle and Run~1 Belle~II datasets, the first search for $B^0 \to K^0_S\tau^+\tau^-$ sets the upper limit on the branching fraction to $\mathcal{B} < 8.3 \cdot 10^{-4}$, the search for $B^+ \to K^+\tau^+\tau^-$ sets $\mathcal{B} < 5.6 \cdot 10^{-4}$, and their isospin combination gives ${\mathcal{B}(B \to K\tau^+\tau^-) < 5.4 \cdot 10^{-4}}$, all at 90\% confidence level (CL$_s$). The first search for inclusive $B \to X_s\nu\bar\nu$ decays at Belle~II sets $\mathcal{B} < 3.3 \cdot 10^{-4}$. Finally, the Belle~II $B^+ \to K^+\nu\bar\nu$ result is reinterpreted in the Weak Effective Theory using its published likelihood. The data prefer an enhanced vector contribution and mildly favour a nonzero tensor contribution.
}

\FullConference{
  43rd International Conference on High Energy Physics (ICHEP2026)\\
  30 July -- 5 August 2026\\
  Natal, Brazil
}

\begin{document}
\maketitle

\section{Introduction}
\label{sec:intro}

Electroweak penguin decays mediated by the flavour-changing neutral-current transitions \mbox{$b \to s\nu\bar\nu$} and \mbox{$b \to s\tau^+\tau^-$} are loop suppressed in the Standard Model (SM), with branching fractions from $\mathcal{O}(10^{-7})$ for $B \to K\tau^+\tau^-$ to $\mathcal{O}(10^{-5})$ for $B \to X_s\nu\bar\nu$~\cite{Parrott:2022zte,Fael:2025xmi}. Belle~II found evidence for $B^+ \to K^+\nu\bar\nu$ decays 2.7 standard deviations above the SM expectation~\cite{Belle-II:2023esi}, prompting interpretations with heavy new physics and light invisible particles~\cite{Altmannshofer:2023hkn,Chen:2024jlj,Gartner:2026clx}. Inclusive $B \to X_s\nu\bar\nu$ decays complement the exclusive modes~\cite{Felkl:2021uxi}. The $b \to s\tau^+\tau^-$ transition probes third-generation couplings, and leptoquarks~\cite{Aebischer:2022oqe} could enhance its rate by up to three orders of magnitude~\cite{Capdevila:2017iqn}. These proceedings present searches for $B^0 \to K^0_S\tau^+\tau^-$ and $B^+ \to K^+\tau^+\tau^-$ with Belle and Belle~II data and their isospin combination, the first search for $B \to X_s\nu\bar\nu$ at Belle~II, and a reinterpretation of the Belle~II $B^+ \to K^+\nu\bar\nu$ measurement.

\section{Belle and Belle~II experiments}
\label{sec:setup}
The Belle detector~\cite{Belle:2000cnh} at the KEKB collider recorded $711~\mathrm{fb}^{-1}$ of $e^+e^-$ collisions at the $\Upsilon(4S)$ resonance. Its successor, Belle~II~\cite{Belle-II:2010dht} at SuperKEKB~\cite{Akai:2018mbz}, recorded $365~\mathrm{fb}^{-1}$ at the $\Upsilon(4S)$ in its first run and has collected $914~\mathrm{fb}^{-1}$ to date~\cite{Belle-II:lumi}. Electroweak penguin decays with neutrinos or $\tau$ leptons in the final state cannot be fully reconstructed and show no distinct features in kinematic distributions. They are nonetheless accessible at Belle and Belle~II because the initial-state four-momentum is known, the branching fraction $\mathcal{B}(\Upsilon(4S) \to B\bar B)$ is larger than 96\%~\cite{ParticleDataGroup:2024cfk}, and the detectors are nearly hermetic. The missing four-momentum of the final-state neutrinos can therefore be inferred.

Reconstructing the non-signal $B$ meson (tagging) constrains the momentum and flavour of the signal $B$ and assigns all remaining tracks and calorimeter deposits to the signal side. In \textit{hadronic tagging}, the non-signal $B$ is fully reconstructed with the Full Event Interpretation (FEI)~\cite{Keck:2018lcd}. This gives the highest purity but low efficiency. \textit{Inclusive tagging} does not reconstruct the non-signal $B$ explicitly and trades purity for efficiency.

\section{Search for $B^0 \to K^0_S \tau^+\tau^-$}
\label{sec:kstautau}

The first-ever search for $B^0 \to K^0_S\tau^+\tau^-$ decays uses the full Belle and Belle~II Run~1 $\Upsilon(4S)$ datasets~\cite{Belle-II:2026bbr}, together with $90~\mathrm{fb}^{-1}$ of Belle and $42~\mathrm{fb}^{-1}$ of Belle~II off-resonance data. Assuming $\mathcal{B}(B^0 \to K^0_S\tau^+\tau^-) = \mathcal{B}(B^0 \to K^0\tau^+\tau^-)/2$, the SM prediction is $\mathcal{B}_{\rm SM} = (0.78 \pm 0.06) \cdot 10^{-7}$~\cite{Parrott:2022zte}. The non-signal $\bar B^0$ is reconstructed with hadronic tagging. The $K^0_S$ is reconstructed in $\pi^+\pi^-$ decays, and both $\tau$ leptons are reconstructed in one-prong decays to a lepton $\ell = e, \mu$, a charged hadron $h$ (mostly $\pi$, $K$), or $\rho(\to\pi\pi^0)$. The reconstructed squared mass of the $\tau^+\tau^-$ system must satisfy $q^2 > 12~\mathrm{GeV}^2$, and $D^-$ decays are vetoed via $M(K^0_S\, h/\rho) \notin (1.8, 1.9)~\mathrm{GeV}$.

The dominant backgrounds are semileptonic $b \to c \to s$ decays of the signal-side $B$. Because these are flavour specific, the non-signal $B$ flavour separates $\ell h$ final states by the lepton charge, giving five categories: $\ell\ell$, $\rho\ell$, $\ell^-h^+$, $\ell^+h^-$, and no-$\ell$. A boosted decision tree (BDT) is trained for each category and experiment, and the branching fraction is extracted from a simultaneous binned maximum-likelihood fit~\cite{Heinrich:2021gyp} to the BDT output in all five categories, across both experiments' datasets. The background model is validated in off-resonance data, $K^0_S$ mass and beam-constrained mass ($M_{\rm bc}$) sidebands, and $B^0 \to D^{(*)-}\ell^+\nu$ decays in the vetoed charm region. 

\begin{figure}[htbp]
  \centering
  \begin{minipage}[b]{0.664\textwidth}
    \centering
    \includegraphics[width=\linewidth]{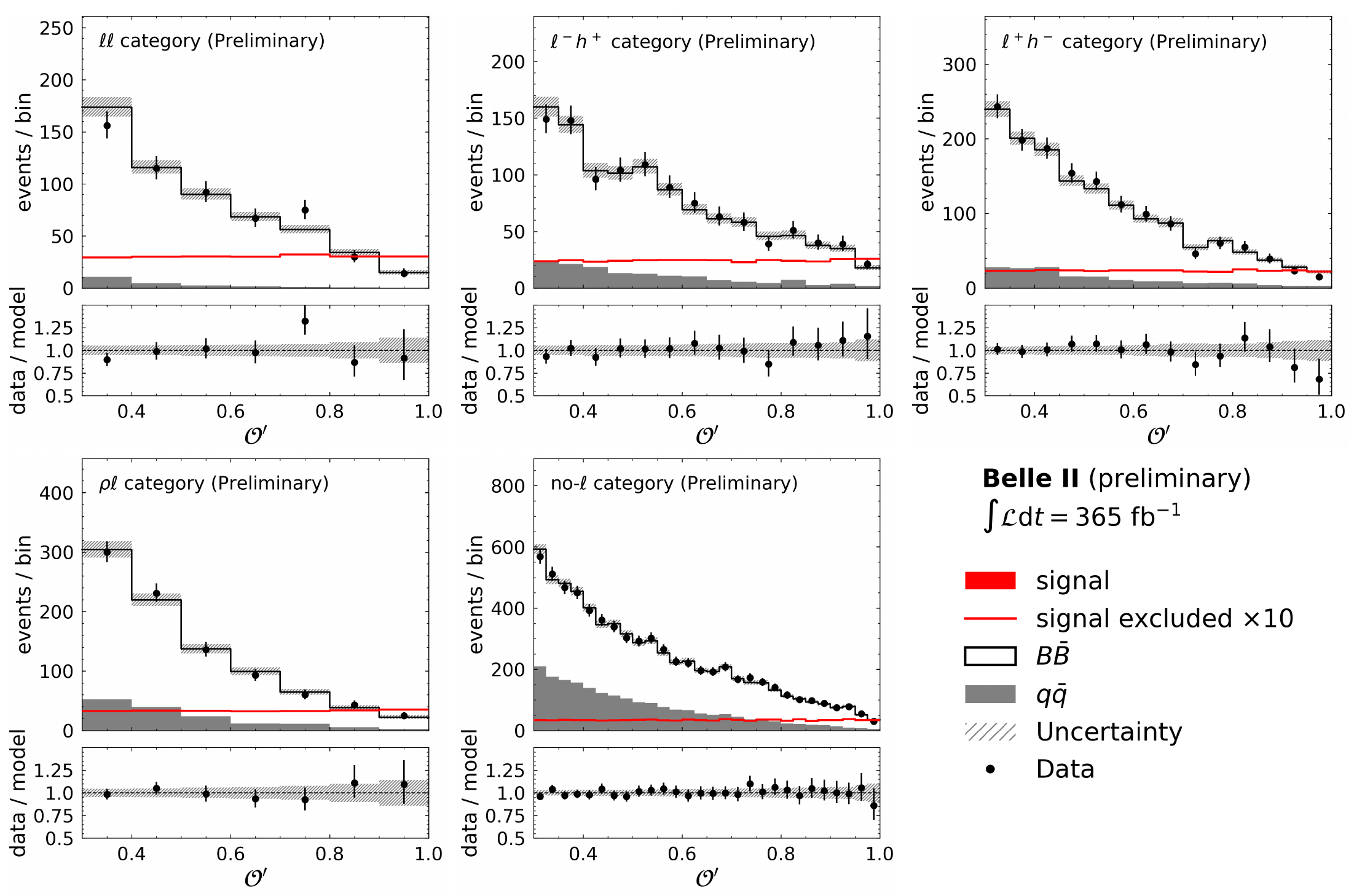}
  \end{minipage}\hfill
  \begin{minipage}[b]{0.296\textwidth}
    \centering
    \includegraphics[width=\linewidth]{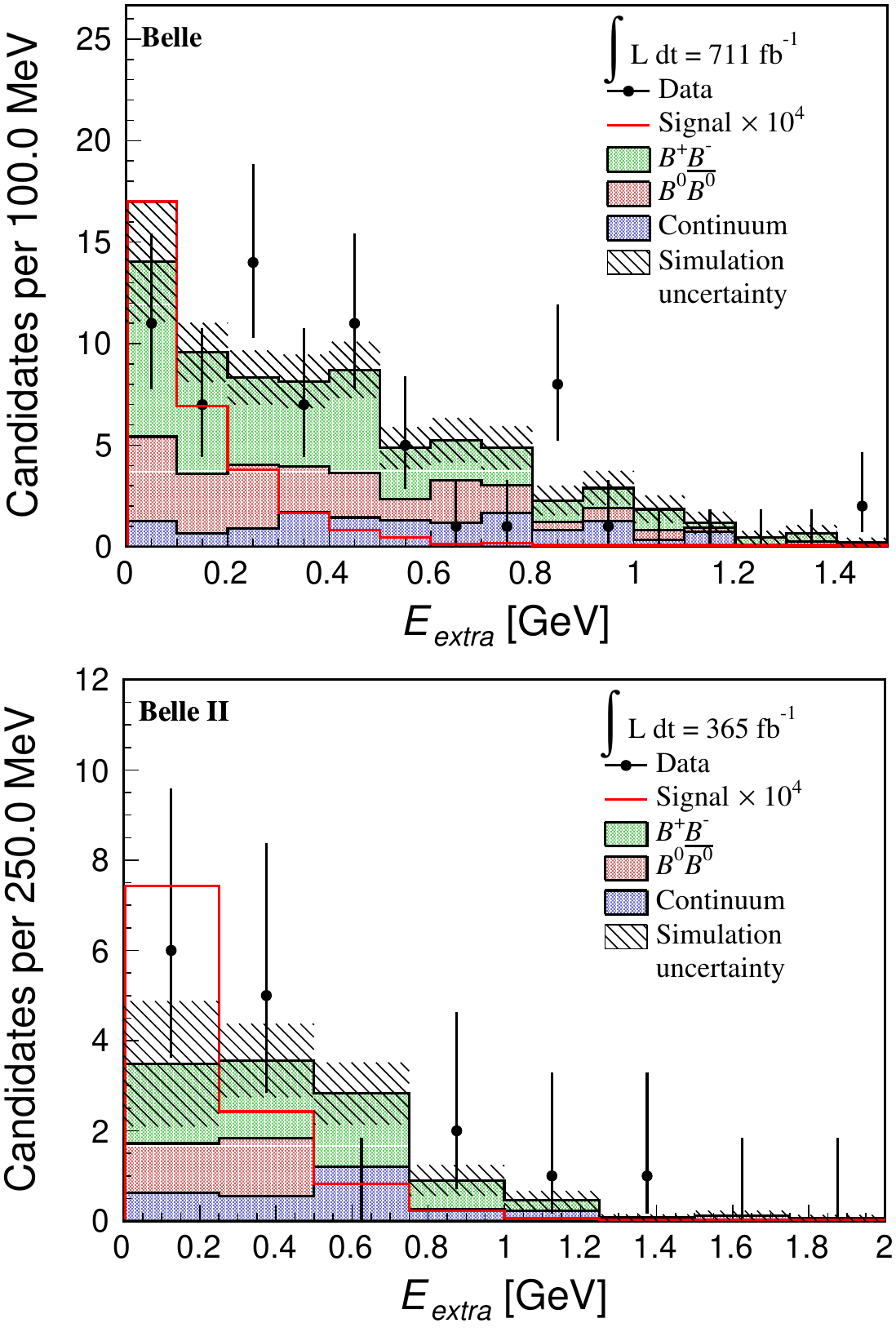}
  \end{minipage}\\
  \begin{minipage}[t]{0.664\textwidth}
    \caption{Post-fit distributions of the transformed BDT output $\mathcal{O}'$
      in the five categories of the Belle~II dataset. The simulated
      $q\bar q$ and $B\bar B$ backgrounds are stacked and compared with data. The 
      red line histogram shows the signal scaled to ten times the upper limit~\cite{Belle-II:2026bbr}.}
    \label{fig:kstautau-postfit}
  \end{minipage}\hfill
  \begin{minipage}[t]{0.296\textwidth}
    \caption{$E_{\rm extra}$ distributions in Belle (top) and Belle~II (bottom)
      data after the final selection, with background expectations from the
      data-corrected simulation~\cite{Belle:2026ooq}.}
    \label{fig:ktautau-eextra}
  \end{minipage}
\end{figure}

The fit gives \mbox{$\mathcal{B}(B^0 \to K^0_S\tau^+\tau^-) = [1.2 \pm 3.3\,(\mathrm{stat}) \pm 2.9\,(\mathrm{syst})] \cdot 10^{-4}$} (Figure~\ref{fig:kstautau-postfit}) and the first-ever upper limit on this decay, $\mathcal{B}(B^0 \to K^0_S\tau^+\tau^-) < 8.3 \cdot 10^{-4}$ at 90\%~CL$_s$~\cite{Belle-II:2026bbr}. The leading systematic uncertainties arise from $B$-decay branching fractions in the background model and the size of the simulated samples.

\section{Search for $B^+ \to K^+ \tau^+\tau^-$}
\label{sec:kptautau}

The search for $B^+ \to K^+\tau^+\tau^-$ decays uses the full Belle and Belle~II Run~1 $\Upsilon(4S)$ datasets~\cite{Belle:2026ooq}. The SM prediction is $\mathcal{B}_{\rm SM} = (1.68 \pm 0.12) \cdot 10^{-7}$~\cite{Parrott:2022zte}. The non-signal $B$ is reconstructed with hadronic tagging. The signal candidate is formed from a kaon with charge opposite to the non-signal $B$ and two opposite-charge leptons ($ee$, $e\mu$, $\mu\mu$) from $\tau^+ \to \ell^+\nu_\ell\bar\nu_\tau$ decays, with $q^2 > 14.18~\mathrm{GeV}^2$ and no additional $\pi^0$ candidates. The dominant $B^+ \to \bar D^{(*)0}\ell^+\nu$ background is suppressed by requiring $M(K^+\ell^-) > 1.9~\mathrm{GeV}$, removing 98.7\% (99.2\%) of it while retaining 20\% (16\%) of the signal in Belle (Belle~II). Further selections on the squared missing mass, the lepton momentum, and the residual calorimeter energy $E_{\rm extra}$ are optimised for the expected upper limit. The signal yield is obtained by counting events in the signal region $E_{\rm extra} < 100~(250)$~MeV for Belle (Belle~II). The background expectation is taken from simulation, corrected by a linear function of $E_{\rm extra}$, fitted simultaneously in three signal-depleted sidebands at low $q^2$, low non-signal-$B$ $M_{\rm bc}$, and high $E_{\rm extra}$.

The Belle dataset yields 11 events (Figure~\ref{fig:ktautau-eextra}) for an expected background of $14.1 \pm 1.6\,(\mathrm{stat}) \pm 1.9\,(\mathrm{syst})$, and the Belle~II dataset yields 6 events for $3.5 \pm 0.7\,(\mathrm{stat}) \pm 0.9\,(\mathrm{syst})$. The combination gives $\mathcal{B}(B^+ \to K^+\tau^+\tau^-) = [0.4\,^{+2.7}_{-2.3}\,(\mathrm{stat})\,^{+1.7}_{-1.8}\,(\mathrm{syst})] \cdot 10^{-4}$ and $\mathcal{B}(B^+ \to K^+\tau^+\tau^-) < 5.6 \cdot 10^{-4}$ at 90\%~CL$_s$, four times more stringent than the only previous bound by BaBar~\cite{BaBar:2016wgb}. The leading systematic uncertainties stem from the estimate of the background yield in the signal region and the shape assumed for the background $E_{\rm extra}$ scaling.

\subsection{Isospin combination}
\label{sec:isospin}
Assuming isospin symmetry, the neutral (Section~\ref{sec:kstautau}) and charged (Section~\ref{sec:kptautau}) modes are combined in a joint likelihood fit with one shared parameter of interest~\cite{Belle-II:2026bbr}
\begin{equation}
  \mathcal{B}(B \to K\tau^+\tau^-) \equiv \mathcal{B}(B^+ \to K^+\tau^+\tau^-)
  = 2\,\frac{\tau_{B^+}}{\tau_{B^0}}\,\mathcal{B}(B^0 \to K^0_S\tau^+\tau^-),
\end{equation}
with the lifetime ratio $\tau_{B^+}/\tau_{B^0} = 1.076 \pm 0.004$~\cite{ParticleDataGroup:2024cfk}. Common systematic uncertainties are correlated. The fit gives \mbox{$\mathcal{B}(B \to K\tau^+\tau^-) = (0.6 \pm 2.8) \cdot 10^{-4}$} and $\mathcal{B}(B \to K\tau^+\tau^-) < 5.4 \cdot 10^{-4}$ at 90\%~CL$_s$.
Adding the neutral mode improves the expected (observed) limit by about 10\%
(5\%) relative to the charged mode alone.

\section{Search for $B \to X_s \nu\bar\nu$}
\label{sec:xsnunu}

\begin{figure}[htbp]
  \centering
  \begin{minipage}[b]{0.52\textwidth}
    \centering
    \includegraphics[width=\linewidth]{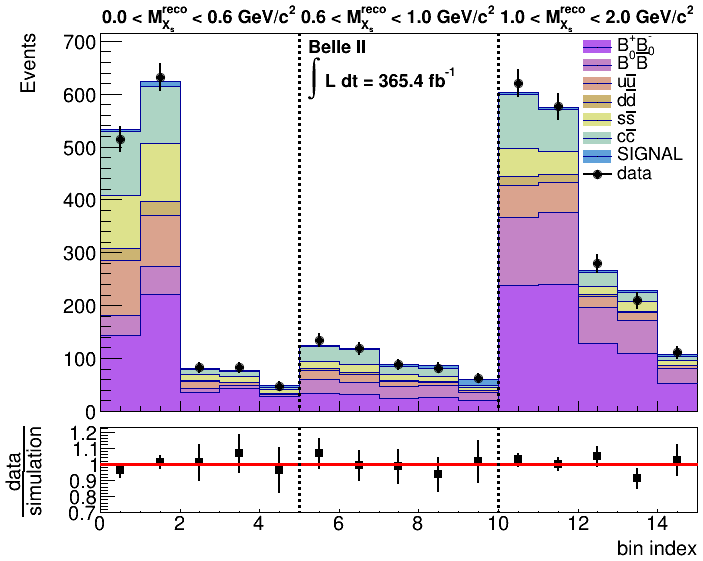}
  \end{minipage}\hfill
  \begin{minipage}[b]{0.44\textwidth}
    \centering
    \includegraphics[width=\linewidth]{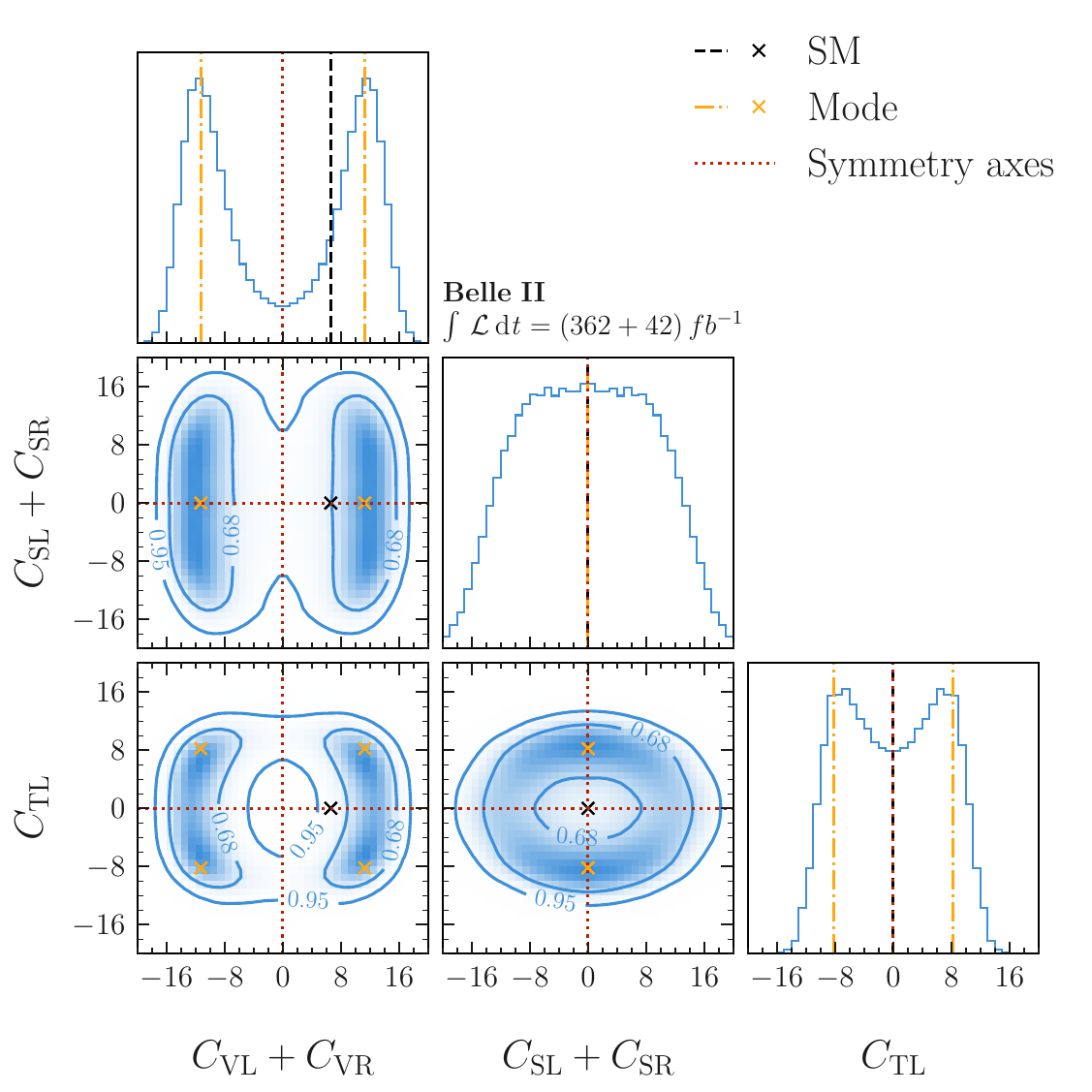}
  \end{minipage}\\
  \begin{minipage}[t]{0.52\textwidth}
    \caption{Post-fit bin-index distribution for data and the fitted signal and
      background components~\cite{Belle-II:2025bho}.}
    \label{fig:xsnunu-postfit}
  \end{minipage}\hfill
  \begin{minipage}[t]{0.44\textwidth}
    \caption{Marginalised posterior for the WET Wilson coefficients~\cite{Belle-II:2025lfq}. Contours indicate the 68\% and 95\%
      credible regions.}
    \label{fig:wet-posterior}
  \end{minipage}
\end{figure}

The first-ever search for inclusive $B \to X_s\nu\bar\nu$ decays uses $365~\mathrm{fb}^{-1}$ of Belle~II $\Upsilon(4S)$ data, together with $43~\mathrm{fb}^{-1}$ of off-resonance data to study the continuum $e^+e^- \to q\bar q$ background~\cite{Belle-II:2025bho}. The SM prediction for the isospin-averaged branching fraction is \mbox{$\mathcal{B}_{\rm SM} = (3.49 \pm 0.11) \cdot 10^{-5}$}~\cite{Fael:2025xmi}. The non-signal $B$ is reconstructed with hadronic tagging, and the $X_s$ system is reconstructed with a sum-of-exclusives method using 30 final states. Backgrounds are suppressed with a BDT, and signal yields are determined with a binned maximum-likelihood fit to the reconstructed $X_s$ mass $M_{X_s}^{\rm reco}$ and BDT output (Figure~\ref{fig:xsnunu-postfit}). Independent signal strengths are extracted in three regions of the true $X_s$ mass, $M_{X_s}^{\rm true} < 0.6$, $0.6$--$1.0$, and $> 1.0~\mathrm{GeV}$. Corrections for continuum modelling, signal fragmentation, photon multiplicity, and major $B$-decay backgrounds are included with their uncertainties. Combining the three $M_{X_s}^{\rm true}$ regions gives \mbox{$\mathcal{B}(B \to X_s\nu\bar\nu) = [8.8^{+8.5}_{-8.2}\,(\mathrm{stat})\,^{+12.6}_{-10.7}\,(\mathrm{syst})] \cdot 10^{-5}$} and $\mathcal{B}(B \to X_s\nu\bar\nu) < 3.3 \cdot 10^{-4}$ at 90\%~CL$_s$~\cite{Belle-II:2025bho}.

\section{Reinterpretation of $B^+ \to K^+ \nu\bar\nu$}
\label{sec:reinterpretation}

The Belle~II evidence for $B^+ \to K^+\nu\bar\nu$ decays uses $362~\mathrm{fb}^{-1}$ of $\Upsilon(4S)$ data, together with $42~\mathrm{fb}^{-1}$ of off-resonance data, and combines hadronic- and inclusive-tag analyses in a joint HistFactory likelihood~\cite{Belle-II:2023esi,Heinrich:2021gyp}. This gives \mbox{$\mathcal{B}(B^+ \to K^+\nu\bar\nu) = [2.3 \pm 0.5\,(\mathrm{stat})\,^{+0.5}_{-0.4}\,(\mathrm{syst})] \cdot 10^{-5}$}, 3.5 standard deviations above zero. 

This result assumes the SM spectrum in the dineutrino squared mass, $q^2$. Since the selection efficiency depends on $q^2$, new physics that changes the spectrum cannot be tested by rescaling the branching fraction~\cite{Gartner:2024muk}. The full likelihood is therefore published together with the joint number density $\nu_{0,x q^2}$ of simulated SM signal events in the reconstructed analysis bins $x$ and the generated (true) $q^2$~\cite{Belle-II:2025lfq,hepdata.166082.v1}. Expected yields for a model predicting the differential branching fraction $\sigma_{1, q^2}$ in each generated $q^2$ bin follow from reweighting~\cite{Gartner:2024muk},
\begin{equation}
  \nu_{1,x} = \sum_{q^2~\mathrm{bins}} \nu_{0,x q^2}\, w_{q^2},
  \qquad
  w_{q^2} = \sigma_{1,q^2} / \sigma_{0,q^2},
\end{equation}
with $\sigma_{0,q^2}$ the SM prediction, as implemented in \textsc{redist}~\cite{redist}. As an application, the result is reinterpreted in the Weak Effective Theory (WET)~\cite{Felkl:2021uxi}, where the differential branching fraction depends on three combinations of Wilson coefficients $C$,
\begin{equation}
  \frac{d\mathcal{B}}{dq^2} =
  \alpha(q^2)\,|C_{\rm VL} + C_{\rm VR}|^2
  + \beta(q^2)\,|C_{\rm SL} + C_{\rm SR}|^2
  + \gamma(q^2)\,|C_{\rm TL}|^2 ,
\end{equation}
where V, S, and T denote vector, scalar, and tensor operators, L and R their chirality, and $\alpha$, $\beta$, and $\gamma$ are determined by the $B \to K$ form factors, evaluated with \textsc{EOS}~\cite{EOSAuthors:2021xpv}. In the SM the only nonzero Wilson coefficient is $C_{\rm VL} = 6.6 \pm 0.1$~\cite{Becirevic:2023aov,EOSAuthors:2021xpv,Parrott:2022zte}. A Bayesian posterior is sampled with uniform priors on these coefficients. The posterior mode is $(|C_{\rm VL} + C_{\rm VR}|, |C_{\rm SL} + C_{\rm SR}|, |C_{\rm TL}|) = (11.3, 0.0, 8.2)$, with 95\% credible intervals $[1.9, 16.2]$, $[0.0, 15.4]$, and $[0.0, 11.2]$~\cite{Belle-II:2025lfq} (Figure~\ref{fig:wet-posterior}). The data prefer an enhanced vector contribution, and the nonzero tensor mode indicates that the SM $q^2$ shape is not the best description of the data. The WET fit has a significance of 3.3 standard deviations over the background-only hypothesis using a likelihood-ratio test statistic.

\section{Summary}
\label{sec:summary}

Belle and Belle~II set the first limit on $B^0 \to K^0_S\tau^+\tau^-$, improve the $B^+ \to K^+\tau^+\tau^-$ limit fourfold, and set \mbox{$\mathcal{B}(B \to K\tau^+\tau^-) < 5.4 \cdot 10^{-4}$} in their isospin combination. The first search for inclusive $B \to X_s\nu\bar\nu$ decays sets \mbox{$\mathcal{B}(B \to X_s\nu\bar\nu) < 3.3 \cdot 10^{-4}$}. The published $B^+ \to K^+\nu\bar\nu$ likelihood allows reinterpretation in new physics models. In the WET, the data prefer an enhanced vector and mildly favour a nonzero tensor contribution. With $914~\mathrm{fb}^{-1}$ already recorded at Belle~II, about twice the Run~1 dataset used here, the sensitivity to these transitions will improve substantially.

\bibliographystyle{unsrt-link}
{\small
\let\oldthebibliography\thebibliography
\renewcommand{\thebibliography}[1]{\oldthebibliography{#1}\setlength{\itemsep}{0pt}\setlength{\parskip}{0pt}}
\bibliography{references}

\begin{thebibliography}{10}

\bibitem{Parrott:2022zte}
W.~G. Parrott, C.~Bouchard, and C.~T.~H. Davies.
\newblock \href{https://doi.org/10.1103/PhysRevD.107.014511}{{Standard Model
  predictions for $B\to K\ell^+\ell^-$, $B\to K\ell_1^-\ell_2^+$ and $B\to
  K\nu\bar\nu$ using form factors from $N_f=2+1+1$ lattice QCD}}.
\newblock {\em Phys. Rev. D}, 107(1):014511, 2023.
\newblock [Erratum: Phys.Rev.D 107, 119903 (2023)].

\bibitem{Fael:2025xmi}
M.~Fael, J.~Jenkins, E.~Lunghi, and Z.~Polonsky.
\newblock \href{https://doi.org/10.1007/JHEP03(2026)217}{{The Standard Model
  prediction for the rare decay $B\to X_s\nu\bar\nu$}}.
\newblock {\em JHEP}, 03:217, 2026.

\bibitem{Belle-II:2023esi}
I.~Adachi et~al.
\newblock \href{https://doi.org/10.1103/PhysRevD.109.112006}{{Evidence for
  $B^+\to K^+\nu\bar\nu$ decays}}.
\newblock {\em Phys. Rev. D}, 109(11):112006, 2024.

\bibitem{Altmannshofer:2023hkn}
W.~Altmannshofer, A.~Crivellin, H.~Haigh, G.~Inguglia, and J.~Martin~Camalich.
\newblock \href{https://doi.org/10.1103/PhysRevD.109.075008}{{Light new physics
  in $B\to K^{(*)}\nu\bar\nu$?}}
\newblock {\em Phys. Rev. D}, 109(7):075008, 2024.

\bibitem{Chen:2024jlj}
F.-Z. Chen, Q.~Wen, and F.~Xu.
\newblock \href{https://doi.org/10.1140/epjc/s10052-024-13425-x}{{Correlating
  $B\rightarrow K^{(*)} \nu \bar{\nu}$ and flavor anomalies in SMEFT}}.
\newblock {\em Eur. Phys. J. C}, 84:1012, 2024.

\bibitem{Gartner:2026clx}
L.~G{\"a}rtner, N.~Krug, T.~Kuhr, M.~A. Schmidt, S.~Stefkova, and B.~Yabsley.
\newblock \href{https://doi.org/10.1103/ggtk-gkx4}{{Constraints on invisible
  $B^+\to K^+X$ decays from the Belle II $B^+\to K^+\nu\bar\nu$ measurement}}.
\newblock {\em Phys. Rev. D}, 114(3):032003, 2026.

\bibitem{Felkl:2021uxi}
T.~Felkl, S.~L. Li, and M.~A. Schmidt.
\newblock \href{https://doi.org/10.1007/JHEP12(2021)118}{{A tale of
  invisibility: constraints on new physics in $b \to s\nu\nu$}}.
\newblock {\em JHEP}, 12:118, 2021.

\bibitem{Aebischer:2022oqe}
J.~Aebischer, G.~Isidori, M.~Pesut, B.~A. Stefanek, and F.~Wilsch.
\newblock \href{https://doi.org/10.1140/epjc/s10052-023-11304-5}{{Confronting
  the vector leptoquark hypothesis with new low- and high-energy data}}.
\newblock {\em Eur. Phys. J. C}, 83(2):153, 2023.

\bibitem{Capdevila:2017iqn}
B.~Capdevila, A.~Crivellin, S.~Descotes-Genon, L.~Hofer, and J.~Matias.
\newblock \href{https://doi.org/10.1103/PhysRevLett.120.181802}{{Searching for
  New Physics with $b\to s\tau^+\tau^-$ processes}}.
\newblock {\em Phys. Rev. Lett.}, 120(18):181802, 2018.

\bibitem{Belle:2000cnh}
A.~Abashian et~al.
\newblock \href{https://doi.org/10.1016/S0168-9002(01)02013-7}{{The Belle
  Detector}}.
\newblock {\em Nucl. Instrum. Meth. A}, 479:117--232, 2002.

\bibitem{Belle-II:2010dht}
T.~Abe et~al.
\newblock \href{https://arxiv.org/abs/1011.0352}{{Belle II Technical Design
  Report}}.
\newblock Technical Report KEK-REPORT-2010-1, arXiv:1011.0352, KEK, 11 2010.

\bibitem{Akai:2018mbz}
K.~Akai, K.~Furukawa, and H.~Koiso.
\newblock \href{https://doi.org/10.1016/j.nima.2018.08.017}{{SuperKEKB
  Collider}}.
\newblock {\em Nucl. Instrum. Meth. A}, 907:188, 2018.

\bibitem{Belle-II:lumi}
{Belle II Collaboration}.
\newblock {Luminosity status}.
\newblock \url{https://www.belle2.org/info/luminosity_status/}, accessed 22
  September 2026.

\bibitem{ParticleDataGroup:2024cfk}
S.~Navas et~al.
\newblock \href{https://doi.org/10.1103/PhysRevD.110.030001}{{Review of
  particle physics}}.
\newblock {\em Phys. Rev. D}, 110(3):030001, 2024.

\bibitem{Keck:2018lcd}
T.~Keck et~al.
\newblock \href{https://doi.org/10.1007/s41781-019-0021-8}{{The Full Event
  Interpretation: An Exclusive Tagging Algorithm for the Belle II Experiment}}.
\newblock {\em Comput. Softw. Big Sci.}, 3(1):6, 2019.

\bibitem{Belle-II:2026bbr}
M.~Abumusabh et~al.
\newblock \href{https://arxiv.org/abs/2607.25607}{{Search for the $B^0 \to
  K^0_{\rm S} \tau^+ \tau^-$ decay}}.
\newblock arXiv:2607.25607 [hep-ex], 7 2026.

\bibitem{Heinrich:2021gyp}
L.~Heinrich, M.~Feickert, G.~Stark, and K.~Cranmer.
\newblock \href{https://doi.org/10.21105/joss.02823}{{pyhf: pure-Python
  implementation of HistFactory statistical models}}.
\newblock {\em J. Open Source Softw.}, 6(58):2823, 2021.

\bibitem{Belle:2026ooq}
M.~Abumusabh et~al.
\newblock \href{https://doi.org/10.1103/12f4-z51q}{{Search for the Decay
  $B^+\to K^+\tau^+\tau^-$ Using Data from the Belle and Belle II
  Experiments}}.
\newblock {\em Phys. Rev. Lett.}, 137(5):051805, 2026.

\bibitem{BaBar:2016wgb}
J.~P. Lees et~al.
\newblock \href{https://doi.org/10.1103/PhysRevLett.118.031802}{{Search for
  $B^{+}\rightarrow K^{+} \tau^{+}\tau^{-}$ at the BaBar experiment}}.
\newblock {\em Phys. Rev. Lett.}, 118:031802, 2017.

\bibitem{Belle-II:2025bho}
M.~Abumusabh et~al.
\newblock \href{https://doi.org/10.1103/kf73-hw61}{{First Search for $B\to
  X_s\nu\bar\nu$ Decays}}.
\newblock {\em Phys. Rev. Lett.}, 136(23):231801, 2026.

\bibitem{Belle-II:2025lfq}
M.~Abumusabh et~al.
\newblock \href{https://doi.org/10.1103/pr66-sd36}{{Model-agnostic likelihood
  for the reinterpretation of the $B^+\to K^+\nu\bar\nu$ measurement at Belle
  II}}.
\newblock {\em Phys. Rev. D}, 112(9):092016, 2025.

\bibitem{Gartner:2024muk}
L.~G{\"a}rtner, N.~Hartmann, L.~Heinrich, M.~Horstmann, T.~Kuhr, M.~Reboud,
  S.~Stefkova, and D.~van Dyk.
\newblock \href{https://doi.org/10.1140/epjc/s10052-024-13038-4}{{Constructing
  model-agnostic likelihoods, a method for the reinterpretation of particle
  physics results}}.
\newblock {\em Eur. Phys. J. C}, 84(7):693, 2024.

\bibitem{hepdata.166082.v1}
\href{https://doi.org/10.17182/hepdata.166082.v1}{{HEPData record for
  ``Model-agnostic likelihood for the reinterpretation of the $B^+\to
  K^+\nu\bar\nu$ measurement at Belle II''}}, 2025.

\bibitem{redist}
L.~G{\"a}rtner.
\newblock \href{https://doi.org/10.5281/zenodo.21780889}{{lorenzennio/redist:
  v2.0.0}}.
\newblock Zenodo, 2026.

\bibitem{EOSAuthors:2021xpv}
D.~van Dyk et~al.
\newblock \href{https://doi.org/10.1140/epjc/s10052-022-10177-4}{{EOS: a
  software for flavor physics phenomenology}}.
\newblock {\em Eur. Phys. J. C}, 82(6):569, 2022.

\bibitem{Becirevic:2023aov}
D.~Be{\v{c}}irevi{\'c}, G.~Piazza, and O.~Sumensari.
\newblock \href{https://doi.org/10.1140/epjc/s10052-023-11388-z}{{Revisiting
  $B\rightarrow K^{(*)} \nu {\bar{\nu }}$ decays in the Standard Model and
  beyond}}.
\newblock {\em Eur. Phys. J. C}, 83(3):252, 2023.

\end{thebibliography}
}

\end{document}